\documentstyle[sprocl,epsf]{article}

\def\Journal#1#2#3#4{{#1} {\bf #2}, #3 (#4)}

\def\be{\begin{equation}}
\def\ee{\end{equation}}
\def\bea{\begin{eqnarray}}
\def\eea{\end{eqnarray}}

\begin{document}

\title{VELOCITY FIELD STATISTICS AND TESSELLATION TECHNIQUES:\\ UNBIASED
ESTIMATORS OF $\Omega$}

\author{Rien van de Weygaert}

\address{Kapteyn Instituut, University of Groningen, P.O. Box 800,\\
9700 AV  Groningen, the Netherlands}

\author{Francis Bernardeau}

\address{Service de Physique Th\'eorique, C.E. de Saclay,\\
F-91191 Gif-sur-Yvette C\'edex, France}

\maketitle\abstracts{We describe two new 
-- {\it stochastic-geometrical} -- methods to obtain reliable velocity 
field statistics from N-body simulations and from any general 
density and velocity fluctuation field sampled at a discrete set of 
locations. These 
methods, the {\it Voronoi tessellation method} and {\it Delaunay 
tessellation method}, are based on the use of the Voronoi and 
Delaunay tessellations of the point distribution defined by the 
locations at which the velocity field is sampled. Adjusting themselves
automatically to the density of sampling points, they represent 
the optimal estimator for volume-averaged quantities. They are 
therefore particularly suited for checking the validity of the 
predictions of quasi-linear analytical density and velocity 
field perturbation theory through the results of N-body simulations 
of structure formation. We illustrate the subsequent 
succesfull application of the two methods to estimate the 
bias-independent value of $\Omega$ in the N-body simulations on 
the basis of the predictions of perturbation theory for the 
$\Omega$-dependence of the moments and PDF of the velocity divergence 
in gravitational instability structure formation scenarios with 
Gaussian initial conditions. We will also shortly discuss practical 
and complicating issues involved in the obvious 
extension of the Voronoi and Delaunay method to the analysis of 
observational samples of galaxy peculiar velocities.}
\vskip 1.0cm

\section{Introduction}
The study of the large-scale cosmic velocity field is a very promising
and crucial area for the understanding of structure formation. The
cosmic velocity field is in particularly interesting because of its 
close relation to the underlying field of mass fluctuations. Indeed, 
on these large and (quasi)-linear scales the acceleration, and 
therefore the velocity, of any object is expected to have an 
exclusively gravitational origin so that it should be independent of 
its nature, whether it concerns a dark matter particle or a bright 
galaxy. Moreover, in the linear regime the generic gravitational 
instability scenario of structure formation predicts that at every 
location in the Universe the local velocity is related to the 
local acceleration, and hence the local mass density fluctuation 
field, through the same universal function of the cosmic density 
parameter $\Omega$ (Peebles 1980), $f(\Omega) \ \propto \ \Omega^{0.6}$.
Because linear theory provides a good description on scales exceeding a 
few Megaparsec, the use of this straightforward relation implies 
the possibility of a simple inversion of the measured velocity field 
into a field that is directly proportional to the field of local 
mass density fluctuations. Such a procedure can then be invoked to 
infer the value of $\Omega$, through a comparison of the resulting 
field with the field of mass density fluctuations in the same region. 

However, such a determination of $\Omega$ may be contrived as the 
estimate of the mass density fluctuation field on the basis of the 
observed galaxy distribution may offer a biased view of the   
underlying mass distribution. By lack of a complete and
self-consistent physical theory of galaxy formation, the commonly
adopted approach is to make the simplifying assumption that 
the galaxy density $\delta_g$ and the mass density $\delta$ are
related via a linear bias factor $b$,
$\delta_g\,=\,b\,\delta$. The comparison between the observed galaxy 
density fluctuation field and the local cosmic velocity field will 
therefore yield an estimate of the ratio $\beta\,=\,{f(\Omega)/b} \,
\approx\, {\Omega^{0.6}/b}\,.$. 
However, while numerous studies have yielded estimates of $\beta$
in the range $\beta \approx 0.5-1.2$ (see Dekel 1994, Strauss \& 
Willick 1995, for compilations of results), it has proven very 
cumbersome to subsequently disentangle the contribution of $\Omega$
and $b$ to the quantity $\beta$. In fact, it turns out to be 
impossible within procedures based on the linearity of the analysed 
velocity field. 

\section{Perturbation theory and the Quasi-linear evolution of the 
velocity field} 
As long as density and velocity fluctuations are small they grow
linear, at a global rate irrespective of the value and location of 
the perturbation. The dependence on the mass of the fluctuation only 
expresses itself through the value of the universal cosmic matter 
density, i.e. through the value of $\Omega$. This specific
circumstance will cease to hold as soon as the fluctuations start to 
acquire values in the order of unity and higher. The larger
gravitational acceleration exerted by the more massive structures in 
the density field then starts to induce the infall of 
proportionally ever larger amounts of matter and hence to an increase of the 
growth rate. This leads to a situation in which the growth of the 
density fluctuation is no longer merely dependent on 
the global value of $\Omega$, but also on the local value of the density 
excess or deficit. This in turn implies that the situation during the
linear regime of the random density and velocity perturbations 
retaining the same statistical properties will no longer
persist. Instead, the nonlinear evolution of the fluctuations leads to
an increasingly asymmetric evolution of the density and velocity 
fluctuations, with overdensities collapsing into compact massive 
objects whose density excess can achieve values exceeding unity by 
many orders of magnitude, while underdensities expand into empty 
troughs whose density deficit is naturally limited to be no lower than
-1.0. 

\begin{figure}
\centering
\epsfxsize=7.5cm
\epsfbox{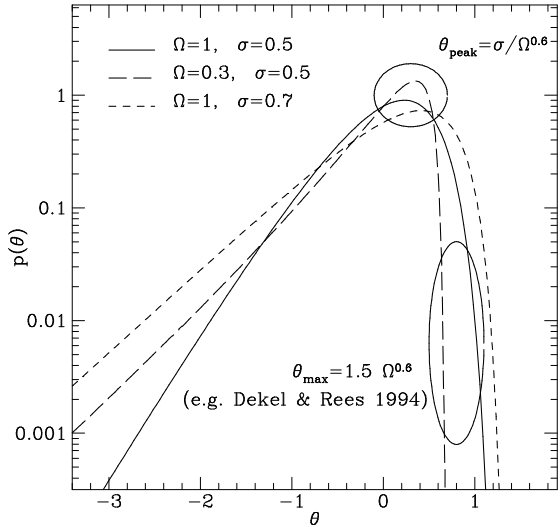}
\caption{The PDF of the velocity divergence $\theta$ for 3 different 
cosmologies, as a function of $\Omega$ and $\sigma_{\theta}$.
\label{fig:PDF}}
\end{figure}

If the primordial random fluctuation field is characterized by
Gaussian statistics, one can apply perturbation theory to 
analytically compute the development of the stochastic properties 
of the field in the first quasi-linear stages as mild nonlinear 
perturbations develop in the random density and velocity field through
the action of gravity (see Juszkiewicz \& Bouchet 1997 for a review 
and references). For the purpose of determining the value 
of $\Omega$, Bernardeau (1994) and Bernardeau et al. (1995) found one 
of the most straightforward and useful results in the context of perturbation 
theory. In an extensive body of work he demonstrated the existence 
of a simple relation between the higher order moments 
$\langle \theta^p \rangle$ of the divergence $\theta$ of the 
locally smoothed velocity field ${\bf v}$, 
\begin{equation}
\theta \,\equiv \,{\displaystyle \nabla \cdot {\bf v} \over \displaystyle 
H}\,,
\label{eq:theta}
\end{equation}
to its second order moment $\langle \theta^2 \rangle$. The coefficient
$T_p$ depends on $\Omega$, on the shape of the power spectrum, the 
geometry of the window function that has been used to filter the 
velocity and even on the value of the cosmological 
constant $\Lambda$. For the case of a {\it top-hat filtered}  
velocity field, analytical expressions for $T_p$ were derived that 
showed them to be strongly dependent on the value of $\Omega$, only 
very weakly dependent on $\Lambda$ and, {\it very significant} within the
present context, completely independent of a linear bias factor
$b$. As in practical circumstances it may still be feasible to get 
reliable estimates of the lower-order moments, this implies that e.g. the 
relation between the skewness and the second-order moment of $\theta$
may be succesfully exploited for the determination of $\Omega$ through the 
coefficient $T_3$,
\begin{equation}
T_3 \equiv \langle \theta^3 \rangle / \langle \theta^2 \rangle^2
\,\propto\,\Omega^{-0.6}\,.
\label{eq:t3}
\end{equation}
In fact, perturbation theory allows to infer the strongly
$\Omega$-dependent, weakly $\Lambda$-dependent, and $b$-independent 
expressions for all coefficients $T_p$, and from this complete series
of coefficients one can construct the complete Probability
Distribution Function $p(\theta)$ (see Bernardeau 1994). An
illustration of the changing 
global shape and behaviour of the complete PDF is evidently more
direct into conveying an impression of the way in which the
statistical properties of the velocity field
depend on the value of $\Omega$ and evolve through gravity. Such an 
illustration is provided by the linear-log plot of $p(\theta)$ for 
3 different cosmologies. Notice that $\Omega$ not only influences the 
overall shape of the PDF but also the location of its peak --
indicated by $\theta_{\rm max}$ -- as well as that of the cutoff at the 
high positive values of $\theta$. The value of the maximum of
$\theta$, at $\theta_{\rm max}$ is directly related to the expansion 
of the deepest voids in such a Universe. 

\section{Volume-averaged quantities}
The perturbation formalism discussed above evidently concerns
continuous density fields. Practical applications to either N-body 
simulations or real observational samples, on the other hand, yield 
velocity fields that are sampled at a finite number of discrete, 
non-uniformly distributed, locations. This {\it discreteness} forms a 
major technical obstacle for the succesfull application of the 
theoretical results. The usual approach is to smooth the 
discrete velocity field by some filter function. Almost without 
exception the conventional filtering schemes concern a {\it
mass-weighted} velocity field filtered with a filter function 
$W_M({\bf x},{\bf x}_0)$ 
\begin{equation}
{\bf v}_{mass}({\bf x}_0) \equiv {\displaystyle \int 
d{\bf x}\,{\bf v}({\bf x}) \,\rho({\bf x}) W_M({\bf x},{\bf x}_0) \over 
\displaystyle \int d{\bf x}\,\rho({\bf x}) W_M({\bf x},{\bf x}_0)}\,.
\label{eq:fmsv}
\end{equation}
An example of this is probably one of the most frequently 
applied class of filtering schemes, involving the interpolation of 
the velocity field values at the random sampling locations to those 
at regular grid locations, weighing the contribution by each sampling 
point by the filter function value. However, the presence of the 
the extra mass-weighting density field factor $\rho$ would introduce 
considerable technical repercussions, and therefore analytical results
within perturbation theory have been almost exclusively limited to 
{\it volume-weighted} filtered velocity fields ${\tilde {\bf v}}$,
\begin{equation}
{\tilde {\bf v}}({\bf x}_0) \equiv {\displaystyle
\,\int d{\bf x}\,{\bf v}({\bf x}) W_V({\bf x},{\bf x}_0) \over 
\displaystyle \int d{\bf x}\,W_V({\bf x},{\bf x}_0)}\,,
\label{eq:fvlv}
\end{equation}
where $W_V({\bf x},{\bf x}_0)$ is the applied weight function. For a
succesfull practical implementation of the analytical results of 
perturbation theory it is then of crucial importance to have reliable 
numerical estimators of volume-averaged quantities. One specific
aspect of such estimators is that from the discrete set of sampled 
velocities they should provide a prescription for the value of the 
velocity field throughout the whole sampling volume, unlike the 
conventional mass-weighted schemes that may restrict themselves 
to estimates at a finite number of positions. 

\begin{figure}
\centering
\epsfxsize=6.0cm
\epsfbox{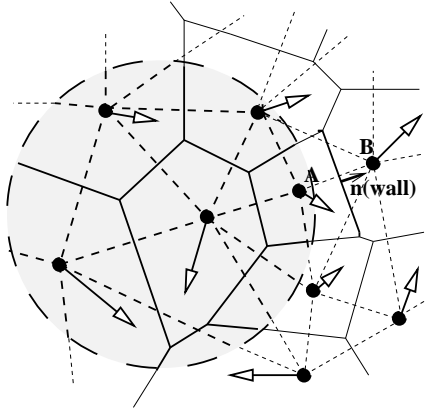}
\caption{Voronoi and Delaunay tessellations of a 2D set of particles
(filled circles). The solid lines form the Voronoi tessellation, the 
dashed lines the Delaunay tessellation. We also indicated a normal
vector {\bf n} of the wall separating the points ${\bf A}$ and ${\bf
B}$. The grey circle represents the area in which one determines 
the top-hat filtered volume average of the velocity gradients. 
\label{fig:vordel}}
\end{figure}

\section{Voronoi and Delaunay Tessellations}
Ideally, the discrete probing and subsequent filtering of the 
velocity field should assign to each location in space 
the value of the velocity at exactly that location. In the infeasible 
situation of being able to do this all over space this would yield 
the continuous velocity field that one attempts to
reconstruct as good as possible from the finite number of discretely 
sampled velocities. In other words, one could consider the underlying 
continuous velocity field as the result of a set of an infinite number
of sampled points volume filtered with a filter function $W_V$ whose 
filter radius is infinitely small. In the case of a
discrete sampling of the field this suggests a velocity field 
reconstruction procedure that defines the velocity at every point 
in space by filtering the corresponding density field -- which can 
be considered as the sum of delta functions peaking at each sample 
location -- with a volume weighted filter that also has a filter 
radius as small as possible. It is then easy to see (see Bernardeau 
\& Van de Weygaert 1996) that this defines a velocity field in 
which the value of the velocity at every location in space acquires the value 
of the velocity at the closest point of the discrete velocity sample. 

The procedure described above implies nothing else than the concept of
the {\it Voronoi tessellation}. Such a tessellation consists of 
a space-filling network of mutually disjunt convex polyhedral cells, 
the {\it Voronoi polyhedra}, each of which delimits the part of space 
that is closer to the defining point in the discrete point sample 
set than to any of the other sample points (see Van de Weygaert 
1991, 1994, for extensive descriptions and references). 

For the application of velocity field analysis, the {\it Voronoi
method} introduced by Bernardeau \& Van de Weygaert (1996) is 
based on the assumption that the velocity field is uniform within 
each Voronoi cell of the tessellation, such that the velocity throughout
each of the Voronoi polyhedra is equal to that of the sample point defining
that cell. This assumption immediately implies that the only non-zero
values of the velocity gradients $\partial v_i / \partial x_j$ are
localized to the (polygonal) Voronoi walls. For the
specific case that the window function for the volume filtering is 
a top-hat filter, the subsequent computation of the volume averages 
of the velocity gradients consists of a relatively simple sum of the 
values of those velocity gradients in each of the walls $k$ intersected or
inside the filter sphere weighted by the surface area $A_k$ of the 
part of the wall located within the sphere. Of special interest to 
us was for instance the locally volume-filtered velocity divergence 
${\tilde \theta}_{\rm Vor}$, which for a radius $R$ of the filter 
sphere is computed from
\begin{equation}
{\tilde \theta}_{\rm Vor}\,=\,{\displaystyle 3 \over \displaystyle 
4 \pi R^3}\,\sum_{{\rm walls}_k}\,{ A_k\,{\bf n}_k\cdot
({\bf v}_{k1}-{\bf v}_{k2})\over  H}\,.
\label{eq:thvor}
\end{equation}
To illustrate the above we refer to the cartoon representation of the 
two-dimensional Voronoi method in Figure 2.

\begin{figure}
\centering
\epsfxsize=9.0cm
\epsfbox{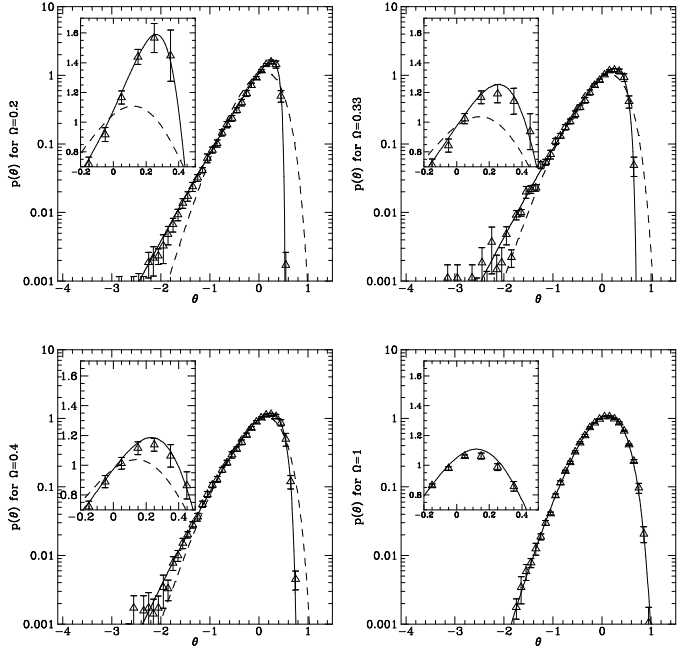}
\caption{The PDF of the velocity divergence for various values of
$\Omega$. The solid lines are the theoretical predictions and the 
dashed lines the predictions for $\Omega=1$ and the same variance. 
The numerical estimates have been obtained using the Delaunay method.
\label{fig:vordelpdf}}
\end{figure}

While the Voronoi method in general leads to good results and in fact 
was succesfully applied to the results of N-body simulations to
confirm the predictions of perturbation theory (see Bernardeau \& 
Van de Weygaert 1996), it evidently represents an artificial situation of 
a discontinuous velocity field. Moreover, its artificial velocity
field implies a few limitations to its application. The most
significant one of these is that it cannot be applied to filter radii 
that are smaller than the average Voronoi wall distance. Below those
scales the probability that a randomly placed filter sphere does not 
contain or intersect any Voronoi wall gets prohibitively high, and 
therefore it would yield unrealistic zero values for the velocity 
gradients. 

It is therefore best to consider the Voronoi method as a zeroth-order 
interpolation scheme, and extend the machinery to include a
first-order interpolation scheme. It is the {\it Delaunay method} that
can be regarded as this elaboration and extension towards a multidimensional
equivalent of a linear interpolation scheme. It is based on the {\it 
Delaunay tessellation} defined by the points in the sample. This
uniquely defined and space-covering network consists of mutually 
disjunt {\it Delaunay tetrahedra}. Each of the Delaunay tetrahedra is 
defined by four nuclei from the point sample that have a 
circumscribing sphere that does not contain any of the other nuclei 
in its interior. In fact, a host of practical, computational and
image-processing, applications already employ the interpolation and 
triangulation qualities of Delaunay tessellations, based on the
realization that they represent the near-optimal multidimensional 
triangulation of a discretely sampled space. For better appreciation, 
we have also illustrated the Delaunay method through its two-dimensional 
equivalent in Figure 2, the dashed lines representing the Delaunay
tessellation. As they are each others dual, the Delaunay and Voronoi
tessellation are closely related. This close relationship is born out 
by the fact that the centre of the circumscribing sphere of a 
Delaunay tetrahedron is a vertex in the corresponding Voronoi 
tessellation. 

The first-order velocity field interpolation scheme defined by our 
{\it Delaunay method} (see Bernardeau \& Van de Weygaert 1996) is 
based on the construction of the velocity field ${\bf
v}(M)$ at every location $M$ in space through linear interpolation 
between the velocities of the four particles $A$, $B$, $C$
and $D$ that define the Delaunay cell in which $M$ is situated.   
${\bf v}(M)\,=\,\alpha_A {\bf v}(A) + \alpha_B {\bf v}(B) + 
\alpha_C {\bf v}(C) + \alpha_D {\bf v}(D)$, 
where the weights $\alpha_{j}$ are the barycentric weights of the 
points. The resultant velocity field is one of a field of uniform velocity 
gradients within each Delaunay cell. The value of each of the nine
velocity gradient components $\partial v_i/\partial v_j$ within each Delaunay 
tetrahedron, and through them the value of the velocity divergence
$\theta$, the shear $\sigma_{ij}$, and even the vorticity
$\omega_{i}$, follow immediately through solving the 9 linear
equations that one obtains through evaluation of the three 
independent velocity differences obtained by evaluating the value of
the velocity at the four different vertex locations of the tetrahedron.
Having defined the values of the velocity gradient, and hence of the 
velocity itself, over the entire sample volume, the last step in 
the Delaunay method then consists of determining the volume averaging 
quantities through top-hat filtering with a filter $W_{TH}$ that has 
a characteristic radius $R$. Essentially this consists of determining 
the weighted average of the value of $\theta$, $\sigma_{ij}$ and/or 
$\omega_i$ in a sphere of radius $R$ centered on a location ${\bf
x}_0$, the weights being the volume of the part of the Delaunay 
tetrahedron that intersects with the filter sphere. 

As the velocity gradients in the Delaunay method will nowhere acquire  
artificial zero values, it is a much more robust method. Moreover, 
it is far less memory consuming to store the information on the 
Delaunay tetrahedra than it is to store the complete geometrical information 
on Voronoi polyhedra, so that it can be applied to truly big 
datasets. The one serious disadvantage in its present state is that 
it is a rather time-consuming method, due to the fact that 
the calculation of the intersection between randomly shaped and 
located tetrahedra and spheres turns out to be anything but a trivial 
problem. 

By applying the procedures sketched above to a large number of
randomly located filter spheres of a particular filter radius $R$, 
one obtains a numerically determined probability distribution of 
the velocity gradients in for example the outcome of N-body 
simulations. In this way one can check the analytical predictions 
of perturbation theory, evidently restricted to the early nonlinear stages 
of such simulations. Moreover, as agreement between the analytical 
predictions and the numerical inferences by the Voronoi and Delaunay 
method provides confidence in both, one can apply the tessellation 
methods to for extending the corresponding statistical study to the highly 
nonlinear structure formation stages whose velocity field
characteristics cannot be adressed by analytical means.  

\section{Results, Discussion and Prospects}
The main incentive for developing the Delaunay and Voronoi method is 
provided by the wish to be able to infer a bias-independent value 
of $\Omega$ through comparison of the velocity statistics obtained 
from the discrete point sample with those of analytical
distributions. In figure 3 we show the PDFs of the velocity divergence
$\theta$ that were numerically determined by the Delaunay 
method for a range of N-body simulations,
each with a different value of $\Omega$. The solid curve shows the 
corresponding analytical distribution function $p(\theta)$, for the 
cosmic epoch with the same dispersion $\sigma_{\theta}$. For contrast,
each of the four frames also contains the dashed curve for the 
PDF in an Einstein-de Sitter universe with the same value of
$\sigma_{\theta}$. Evidently, the Delaunay method is highly succesfull
in reproducing 
the correct statistical distribution, and via the relations between 
the moments of the PDF we indeed obtain very good estimates of 
$\Omega$. 

While figure 3 illustrates the potential power of the tessellation 
methods, we are obviously motivated to apply them to more practical 
situations and hence more cumbersome cases where selection and 
sampling effects and sampling errors are of crucial
influence. In particular we hope to be able to develop a formalism 
capable of dealing with observational catalogues of peculiar 
velocities of galaxies. In previous work (Bernardeau \& Van de
Weygaert 1996, Bernardeau et al. 1997) we already adressed the 
issue of diluted samples. In those cases both methods yielded 
encouraging results. 
However, the true world will present problems ranging from the 
fact that one can measure galaxy velocities only along the line 
of sight to complicated selection effects like differential 
Malmquist bias (see e.g. Bertschinger et al. 1990, 
Dekel, Bertschinger \& Faber 1990). Work on 
these issues is in progress, but they obviously provide a considerable 
complication. 

\section*{Acknowledgments}
We wish to thank Eric Hivon and Fran{\c c}ois Bouchet, our collaborators in
part of the project, and Volker M\"uller for the 
excellent organization of this Potsdam meeting.


\section*{References}
\begin{thebibliography}{99}
\bibitem{br1}F. Bernardeau, \Journal{ApJ}{292}{1}{1992}

\bibitem{bjdb}F. Bernardeau, R. Juszkiewicz, A. Dekel and F. Bouchet, 
\Journal{MNRAS}{274}{20}{1995}

\bibitem{bw1}F. Bernardeau and R. van de Weygaert, \Journal{MNRAS}{279}{693}{1996}

\bibitem{bwhb}F. Bernardeau, R. van de Weygaert, E. Hivon and
F. Bouchet, MNRAS, 1997, in press

\bibitem{bdfdb}E. Bertschinger, A. Dekel, S.M. Faber, A. Dressler and 
D. Burstein, \Journal{ApJ}{364}{370}{1990}

\bibitem{d1}A. Dekel, \Journal{ARAA}{32}{371}{1994}

\bibitem{dbf}A. Dekel, E. Bertschinger and S.M. Faber, \Journal{ApJ}{364}{349}{1990}

\bibitem{jb}R. Juszkiewicz and F. Bouchet, manuscript, 1997

\bibitem{sw}M.A. Strauss and J.A. Willick, \Journal{Physics
Rep.}{261}{271}{1995}

\bibitem{w1}R. van de Weygaert, Ph.D. thesis, Leiden University, 1991

\bibitem{w2}R. van de Weygaert, \Journal{A\&A}{283}{361}{1994}

\end{thebibliography}
\end{document}